\documentclass[fleqn,10pt]{wlscirep}
\usepackage[utf8]{inputenc}
\usepackage[T1]{fontenc}
\usepackage[squaren]{SIunits}
\usepackage{tikz}
\usepackage[percent]{overpic}

\usepackage{float}
\usepackage{subcaption}
\title{Low-dose, high-resolution CT of infant-sized lungs via propagation-based phase contrast}

\author[1,*]{ James A. Pollock}
\author[1]{Kaye Morgan}
\author[1]{Linda C. P. Croton}
\author[2, 3]{Emily J. Pryor}
\author[2, 3]{Kelly J. Crossley}
\author[4]{Christopher J. Hall}
\author[4]{Daniel H{\"a}usermann}
\author[4]{Anton Maksimenko}
\author[2, 3]{Stuart B. Hooper}
\author[1]{Marcus J. Kitchen}

\affil[1]{School of Physics and Astronomy, Monash University, Clayton, VIC, Australia}
\affil[2]{Ritchie Centre, Hudson Institute of Medical Research, Clayton, VIC, Australia}
\affil[3]{Department of Obstetrics and Gynecology, Monash University, Clayton, VIC, Australia}
\affil[4]{ANSTO, based at the Imaging and Medical Beamline of the Australian Synchrotron, Clayton, VIC, Australia }
\affil[*]{Corresponding author; james.pollock@monash.edu}


\begin{abstract}
Many lung diseases require high-resolution imaging for accurate diagnosis and treatment. Computed tomography (CT) is the gold-standard technique for non-invasive lung disease detection, but it presents a risk to the patient through the relatively high ionising radiation dose required. Utilising the X-ray phase information has demonstrated improvements in image quality over absorption contrast in small animal models, at equal or lower radiation levels. Propagation-based phase-contrast imaging requires only a spatially coherent wavefield and some propagation between the sample and detector, making it well suited for medical applications. In particular, lung imaging significantly benefits from the strong phase gradients introduced by the lung-air material interfaces. Herein, propagation-based phase contrast CT is demonstrated stepping up to large animals, namely lambs, as a model for paediatric patients, using monochromatic radiation and a photon-counting detector. The resulting CT images demonstrate superior resolution to existing high-resolution CT systems, and push dose to the quantum limit to comply with current Australian guidelines for infant chest CT exposure of $<2.5\:\text{mSv}$ effective dose. Constituent raw projections are shown to have significant proportions of pixels with zero photon counts that would create severe information loss in conventional CT. Phase retrieval enabled clear visualisation of minor lung airways (\unit{\ge 290}{\mu m}) at doses up to 1,225$\pm$31\% times lower than conventional CT reconstruction and a voxel size of just \unit{75}{\mu m}.
\end{abstract}
\begin{document}

\flushbottom
\maketitle
\thispagestyle{empty}

\section{Introduction}
\label{sec:Introduction}
Phase-contrast imaging has consistently been shown to provide better image quality than absorption-based imaging, particularly for low-dose soft tissue imaging, which is of greatest significance for medical imaging applications \cite{kitchen_ct_2017, albers_high_2023, arhatari_x-ray_2021}. In particular, propagation-based X-ray phase-contrast imaging (PBI) requires minimal adaptation of existing clinical systems and readily combines with computed tomography (CT) to provide complete 3D reconstructions of sample volumes. While clinical CT necessitates higher doses than single image radiography, the incredible stability of PBI in high-noise situations allows for significant dose reduction \cite{kitchen_ct_2017, paganin_simultaneous_2002}, mitigating concerns of patient safety. These improvements may allow for the safe observation of fine structures beyond the ability of conventional absorption-based CT \cite{guo_calibration_2021}, avoiding the need for invasive histology \cite{park_mortality_2007, mazzone_bronchoscopy_2002} and improving clinical imaging systems and patient outcomes.

The benefits of PBI CT have been observed in ex vivo imaging of breast tissue \cite{gureyev_investigation_2014, nesterets_feasibility_2015}. Recent studies have explored energy optimisations \cite{oliva_experimental_2020}, including using radiological assessment via visual grading studies \cite{wan_effect_2021}. Comparisons to conventional absorption imaging have also been performed \cite{gunaseelan_propagation-based_2023, brombal_image_2019}, showing favourable results. Investigations of other biological tissues have progressed through small animal studies, such as the polychromatic source imaging of excised rat hearts \cite{lioliou_framework_2024}. In situ (directly in place) brain imaging of rabbit kittens, using synchrotron radiation, has also been demonstrated to greatly enhance feature visibility \cite{croton_situ_2018, beltran_interface-specific_2011}, although the benefits achievable through PBI imaging have been partially limited by the weaker x-ray phase shifts seen from the bone-brain material combination \cite{beltran_2d_2010}. While this may be negated by alternative analysis techniques \cite{ullherr_correcting_2015, pollock_robust_2023}, research is ongoing to minimize artefacts arising due to the highly-attenuating bone \cite{croton_imaging_2018}. In contrast, the soft-tissue/air interface provides the largest refractive index gradient in the mammalian body, hence the many air/tissue boundaries in the lung create very strong phase contrast effects \cite{murrie_live_2015, saccomano_synchrotron_2018, yagi_refraction-enhanced_1999}. In particular, Kitchen et. al demonstrated the capability of lung PBI to enable large dose reduction factors whilst preserving image quality \cite{kitchen_ct_2017}. This potential has been further explored by Albers et. al using excised pig lungs contained within an artificial co-polymer housing \cite{albers_high_2023}, where clinically relevant dose levels were achieved using an x-ray energy of \unit{40}{keV}, the highest energy currently available at the SYRMEP beamline of the Elettra synchrotron. Herein we present the first in situ phase-contrast CT study of large animal lungs at the Imaging and Medical Beamline (IMBL) of the Australian Synchrotron. Specifically, we imaged the chests of newborn lambs as a model for pediatric lung imaging, optimising the x-ray energy and sample-to-detector distance to achieve high-resolution images for the lowest possible radiation dose afforded by phase contrast. Lung imaging is particularly important for assessing aeration at birth as the lungs may not be fully cleared of the liquid that fills them in utero, and such cases often require ventilatory support \cite{hooper_respiratory_2016}. High-resolution CT can provide accurate assessment of regional lung air volumes, but tissues in infants are also most sensitive to radiation; hence, dose minimisation is essential.

In addition to the applications in paediatrics, clinical chest CTs have also been employed for diagnosing various conditions such as emphysema and chronic obstructive pulmonary disease (COPD) \cite{labaki_role_2017, lynch_quantitative_2013, ostridge_present_2016}, cystic fibrosis \cite{ciet_chest_2023, bortoluzzi_impact_2020}, and cancer \cite{silva_low-dose_2022, lanni_early_2018}, where early diagnosis can be a critical factor for patient prognosis \cite{kauczor_functional_2004}. Although higher resolution imaging aids in diagnosis, dose requirements in CT scale inversely against the fourth power of the voxel size \cite{simonov_restrictions_2004}, creating concerns for patient welfare. This further promotes the need to explore new techniques to improve image resolution, for which PBI combined with phase retrieval offers great promise \cite{kitchen_ct_2017}. 

\begin{figure}[tb!]
	\centering\hfil
	\begin{overpic}[width=0.30\columnwidth,]{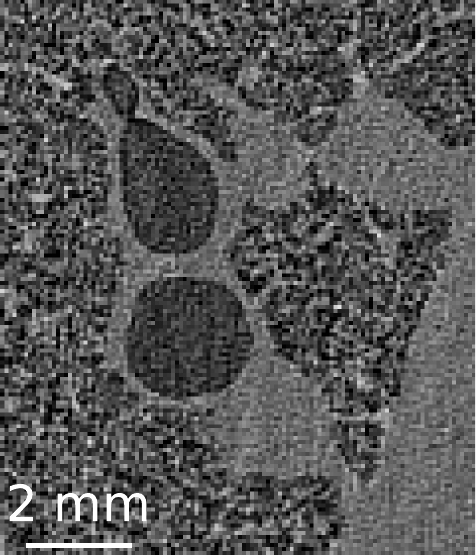}
    \put(72.5,7.5){\shortstack[l]{\fontsize{22}{24}\selectfont \textcolor{white}{a)}}}
    \end{overpic}\hspace{0.5em}
     \begin{overpic}[width=0.3\columnwidth,]{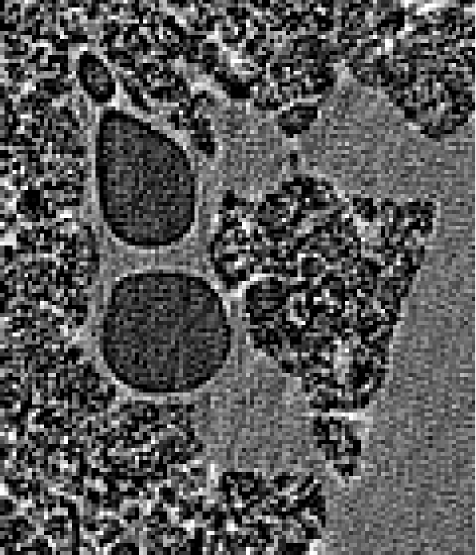}
    \put(72.5,7.5){\shortstack[l]{\fontsize{22}{24}\selectfont \textcolor{white}{b)}}}
    \end{overpic}\hspace{0.5em}
    \begin{overpic}[width=0.3\columnwidth,]{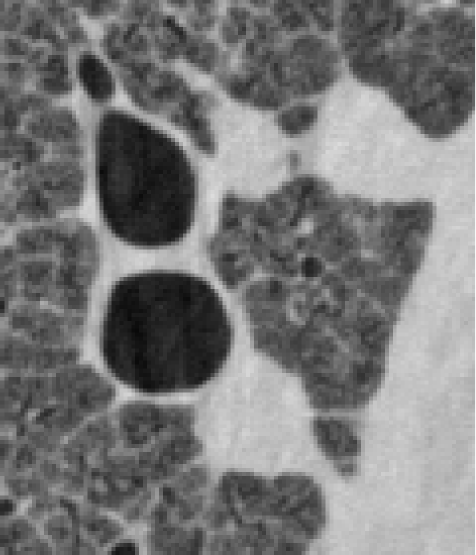}
    \put(72.5,7.5){\shortstack[l]{\fontsize{22}{24}\selectfont \textcolor{white}{c)}}}
    \end{overpic}
    \hfil
    \caption{Phase contrast effects on CT slices demonstrated using aerated lungs of newborn lambs (post-mortem). (a) shows a phase contrast slice recorded at \unit{1}{m} propagation, hence depicting minimal phase contrast fringes in comparison to (b) recorded at \unit{4}{m}. (c) applies phase retrieval to (b), removing the contrast fringes while greatly reducing the noise. }
    \label{fig: PC and PR demo}
\end{figure}

Propagation-based phase contrast imaging is a simple imaging technique requiring a coherent X-ray wavefield, achievable with micro-focus sources \cite{wilkins_phase-contrast_1996, lioliou_framework_2024}, and sufficient distance between the sample and detector for beam propagation and self-interference, typically of the order of a meter. X-rays refracted at boundaries between media propagate to interfere with non-refracted waves, creating phase contrast fringes. In CT, these fringes highlight and sharpen material boundaries, as demonstrated in Figures \ref{fig: PC and PR demo}(a) and (b). These boundaries can be restored to their original resolution through the application of phase retrieval. When considering relatively small propagation distances, such that contrast fringes only contain a single Fresnel fringe pair, the Transport of Intensity Equation (TIE) may be used to derive a highly stable, single-image phase retrieval algorithm \cite{paganin_simultaneous_2002}. Here, phase retrieval is performed using a specifically tuned low-pass filter that suppresses noise in the process of restoring phase contrast fringes (Figure \ref{fig: PC and PR demo}(c)).  
For single-material objects under plane wave illumination TIE phase-retrieval is evaluated through
\begin{equation}
I(x, y) = \mathcal{F}^{-1}\left\{\frac{\mathcal{F}[I_d(x, y)/I_0]}{1 + \frac{\delta(E) \Delta}{\mu(E)}\textbf{k}_{\perp}^2}\right\} \label{equation: paganin},
\end{equation}
where $\mathcal{F}$ and $\mathcal{F}^{-1}$ are the forward and inverse Fourier transform operators, $I_d(x, y)$ is the image measured at the detector plane, $I_0$ is the spatially variable incident intensity, $\Delta$ is the propagation distance from the object to the detector, and $\textbf{k}_{\perp}^2$ describes the transverse spatial frequency components, resulting in a specialised low-pass Lorentzian filter. The real part of the refractive index decrement is given by $\delta$, and $\mu$ is the linear attenuation coefficient, both of which are energy-dependent properties of the sample material. From equation \ref{equation: paganin}, distance and energy are the key parameters that can be adjusted to change both the collected image and the required phase retrieval parameters, and hence can be optimised. This manuscript presents such an optimisation for high-resolution in situ imaging of lamb lungs, which are of similar size to those of human infants. Photon-counting detectors, possessing negligible image blurring and no electronic noise, are incorporated to improve resolution over conventional indirect detectors\cite{danielsson_photon-counting_2021} while further facilitating dose reduction. Additionally, the effects of noise on the image spatial resolution is thoroughly evaluated using Fourier ring correlation. Results demonstrate the viability of phase retrieval in increasing CT resolution whilst retaining safe dose limits, progressing work toward implementation using polychromatic sources and eventually in clinical settings.   

\section{Methods}
Experiments were performed at the Imaging and Medical Beamline (IMBL) of the Australian Synchrotron \cite{stevenson_quantitative_2017} using the wiggler insertion device at \unit{1.4}{T}. The sample was positioned approximately \unit{137}{m} from the source, leading to an average magnification of 1.03 (4m propagation) which was considered negligible during analysis. We first conducted image quality optimisation, focusing on beam energy and propagation distance, then investigated how low the radiation dose could be lowered. All data was acquired using an Eiger 2M-W photon-counting detector (DECTRIS AG, Switzerland) with a \unit{75}{\mu m} pixel size. Imaging parameters that were kept constant in all experiments included the \unit{30}{ms} exposure time and the 0.1 degree inter-frame angle, resulting in scans of 1800 projections taking approximately 55 seconds. The beam was incident on the sample throughout the scan, and time between projections was \unit{9}{\micro s}, comprising less than 0.03\% of the exposure time, and hence did not significantly contribute to the radiation dose delivered. Lambs used during imaging were collected from unrelated, terminal experiments in the same instance as Pryor \textit{et al.} \cite{pryor_quantifying_2022}. Experiments were performed roughly 12 hours after euthanasia via an overdose of sodium pentobarbitone administered immediately after pre-term delivery at approximately 139 days gestation. Animals were surgically intubated post-mortem and the lungs inflated with nitrogen before the trachea were clamped. Nitrogen was used, due to its decreased permeability across the lung barrier, to minimise collapse of the lungs during scanning. The lambs were scanned up-right in a half-cylinder (PMMA) sample holder affixed to the rotation stage. They were held in place by a wire looped behind the teeth to support the head while their forelimbs and lower abdomen were further secured by bandages to limit motion artifacts   

Optimisation datasets were collected in two parts, one to determine the sample-to-detector propagation distance and the other for determining the X-ray energy. Propagation distances were tested at \unit{45}{keV} at 1, 2, 3, 4, 5, and \unit{6}{m} using fixed X-ray flux to the object, while energy optimisations datasets were recorded at 40, 45, 50, and \unit{55}{keV} using a \unit{3}{m} propagation distance. Incident synchrotron radiation was filtered using the beamline's in-vacuum filters of carbon (\unit{0.45}{mm}) and aluminium (\unit{2}{mm}) plates and post-vacuum using \unit{10}{mm} of aluminium. Images were reconstructed at the beamline, with results visually inspected to select optimums for conducting the low-dose experiments. Image quality parameters were properly quantified after the beamtime.  

Low-dose datasets were recorded at \unit{4}{m} propagation distance, using energies of 40, 50, and \unit{60}{keV} to explore the associated effect on spatial resolution. An additional \unit{20}{mm} of post-vacuum aluminium plates were introduced to decrease flux to low-dose levels, totalling \unit{30}{mm}. Flux was further reduced incrementally by adjusting the monochromator detuning; however, this was found to introduce significant energy harmonics. Hence, flux and dose were instead altered by using the roll-off of the synchrotron beam, as the X-ray beam intensity varies along the height of the detector, and hence across different vertical slices of the CT stack.

\subsection{Figures of Merit}
\label{sec: Figures of merit}
Comparing images requires metrics to quantify the desired image qualities, such as the signal-to-noise ratio (SNR). Normalising by radiation dose, to which SNR has a square root proportionality, emphasises patient safety during optimisation. Sections \ref{sec: Experimental optimisation} and \ref{sec: Low dose studies} follow this metric, using the attenuation coefficient of lung tissue as the contrast signal
\begin{align}
    \text{SNR}/\sqrt{\text{Dose}} &= \frac{\Bar{\mu}_{tissue}}{\sigma_{tissue} \sqrt{D}}, \label{eq: SNRperDose}
\end{align}
where $D$ is the dose measured, $\Bar{\mu}_{tissue}$ is the mean value of tissue voxels measured within a CT slice, and $\sigma_{tissue}$ is the standard deviation. Voxels containing tissue were selected using manual material segmentation (such as that shown in Figure \ref{fig: Dose example}) with erosion filters applied to ensure only tissue voxels were measured. Spatial resolution ($R$) is a key image quality parameter, which may be incorporated according to the intrinsic image quality factor ($Q$)
\begin{align}
    Q &= \frac{\Bar{\mu}_{tissue}}{\sigma_{tissue} \sqrt{D R^{3}} }. \label{eq: SNR/dose/res}
\end{align}
where resolution is raised to the third power to reflect the measurements being performed in 3D \cite{gureyev_spatial_2016}. Note, however, that resolution measurements are often a fixed value for a given detector system, measured as the system blurring width using high flux, which is then applied to all subsequent measurements. However, this approach disregards the effect of noise on spatial resolution, an issue which dominates at low flux. Resolution is arguably better measured as the minimal feature length that can be consistently resolved in the presence of noise. Consequently, measurements of $Q$ in Section \ref{sec: Low dose studies} employ the noise-limited resolution metric found using Fourier Ring Correlation (FRC$_{\text{or}}$), as detailed in \cite{van_heel_fourier_2005}. All SNR and dose measurements were taken over a range of 10 CT slices, with the resulting standard deviation between individual slice measurements as the uncertainty in each measurement. Resolution measurements from the FRC$_{\text{or}}$ were also taken over the same slice range; however, the standard error is used to reflect the uncertainty of an averaged measurement. 

\subsection{Quantifying Spatial Resolution}
\label{sec:FRC}
Spatial resolution represents the smallest feature size capable of being resolved in a given detector system. These measurements commonly quantify blurring effects attributed by various system elements, such as the X-ray source size and detector point spread function (PSF), and are typically taken where the effect of noise is minimal. Developments in photon-counting detectors now allow imaging with near single-pixel PSFs \cite{fardin_characterization_2023}, meaning that noise effects become the dominant influence on spatial resolution in the low-dose regime. In this context FRC$_{\text{or}}$, which quantifies the minimum feature size that can be resolved from noise, becomes a more appropriate metric. FRC$_{\text{or}}$ is performed by quantifying the SNR measured within Fourier frequency bins. FRC$_{\text{or}}$ is evaluated using two independently recorded images, $f_1(r)$ and $f_2(r)$, through 
\begin{align}
    \text{FRC}_{\text{or}}(r) &= \frac{\sum_{r\in r_i} F_1(r) F_2(r)^{*}}{\sqrt{\sum_{r\in r_i} F_1^2(r)\sum_{r\in r_i} F_2^2(r)}}, \label{eq:FRC}
\end{align}
where capitalisation denotes the Fourier-space image, $^*$ the complex conjugate, and $\sum_{r\in r_i}$ represents the sum over all Fourier space voxels (r) contained within the spatial frequency shell $(r_i)$. $\text{FRC}_{\text{or}}(r_i)$ is then the correlation between $f_1(r)$ and $f_2(r)$ in the frequency range $r_i$. Images must be recorded such that their common signal may be compared to the independently-sampled noise fluctuations, allowing the SNR to be deduced. For CT, this can be done by recording two separate datasets, using precisely aligned images, or by splitting a single recording into two half-datasets. FRC$_{\text{or}}$ analysis in this paper is performed with the latter method
by making two CT reconstructions from the full dataset using alternating projections but beginning with a common projection to ensure alignment. 
\begin{figure}[b!]
    \centering
    \begin{overpic}[width=0.70\textwidth,,]{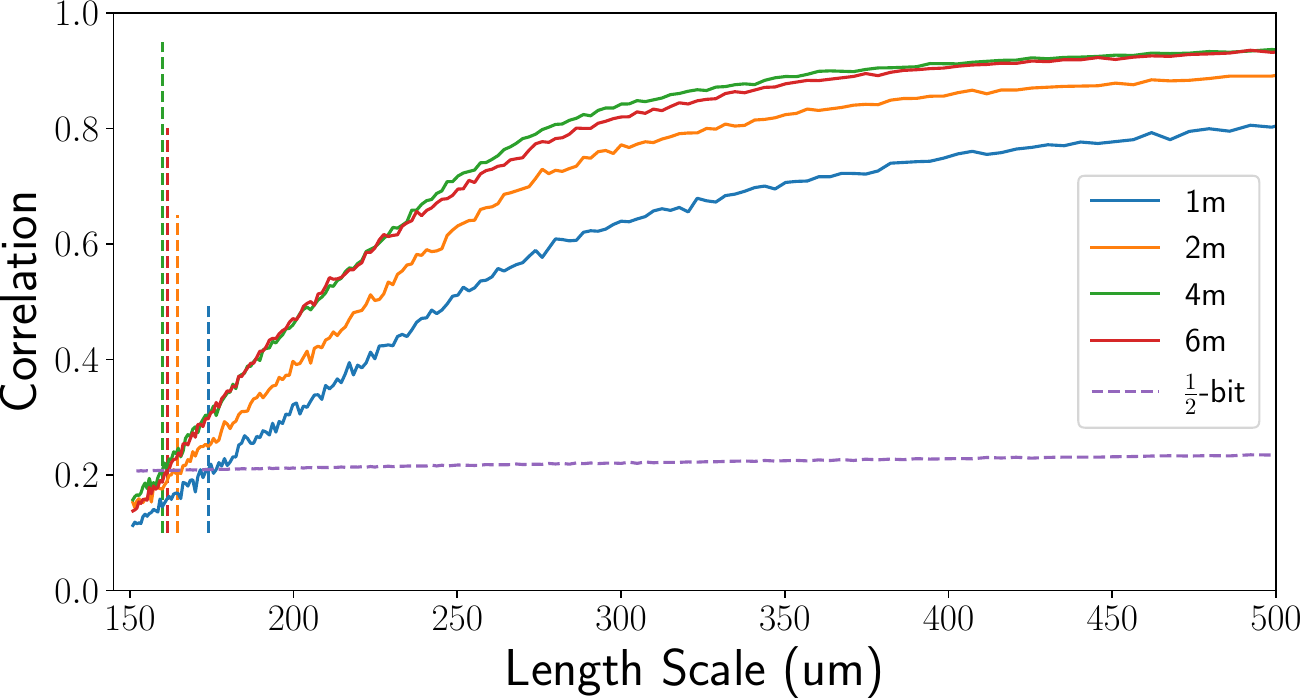}
    \put(15, 31.5){\shortstack[l]{\fontsize{8}{18}\selectfont \textcolor{black}{\unit{174}{\mu m}}}}
    \put(15, 37.0){\shortstack[l]{\fontsize{8}{18}\selectfont \textcolor{black}{\unit{164}{\mu m}}}}
    \put(15, 43){\shortstack[l]{\fontsize{8}{18}\selectfont \textcolor{black}{\unit{161   }{\mu m}}}}
    \put(15, 48){\shortstack[l]{\fontsize{8}{18}\selectfont \textcolor{black}{\unit{160}{\mu m}}}}
    \end{overpic}
    \caption{FRC$_{\text{or}}$ trends of phase contrast CTs (without phase retrieval, as seen in Figure \ref{fig: PC and PR demo}(b)), focusing on the smallest length scales. All trends were recorded at 45 keV and propagation distances of 1, 2, 4, and 6m, demonstrating increasing correlation due to phase contrast that plateaus beyond 4m. Vertical dashed lines denote the intersections between the FRC$_{\text{or}}$ trends and the half-bit threshold used to define resolution. Datasets were collected using the Eiger 2M-W photon counting detector and here contain a maximum length scale of \unit{60}{mm}.}
	\label{fig: PC FRC}
\end{figure}

The measurement $\text{FRC}_\text{or}(r)$ tends toward 1 where signal dominates noise and tends towards zero where noise is dominant. Important consequences of Equation \eqref{eq:FRC} are that multiplicative factors divide out as they preserve the SNR ratio. This means that TIE-based phase retrieval, under ideal conditions, will not affect the FRC$_{\text{or}}$ since it can be represented by divisors applied to each Fourier frequency bin. However the effect of phase contrast, by highlighting features of the object and not noise, enhances signal components in the Fourier transforms, and hence increases correlation between images. 

The spatial resolution is determined as the spatial length at which the $\text{FRC}_{\text{or}}(r)$ intercepts some threshold. While previous studies define static correlation thresholds, such as a 0.5 or 1/7, the proceeding analysis follows \cite{van_heel_fourier_2005} by defining a varying threshold, $T(r_i)$, that directly relates to the SNR in each frequency bin as
\begin{equation}
    T(r_i) = \frac{\text{SNR}(r_i) + 2\sqrt{\text{SNR}(r_i)}/\sqrt{n(r_i)} + 1/\sqrt{n(r_i)}}{\text{SNR}(r_i) + 2\sqrt{\text{SNR}(r_i)}/\sqrt{n(r_i)} + 1}. \label{eq: SNR threshold}
\end{equation}
Equation \ref{eq: SNR threshold} allows specific SNR cutoffs to be set for the resolution, such as the 1-bit threshold
\begin{align}
    T_{\text{1-bit}}(r_i) &= \frac{0.5 + 2.4142/\sqrt{n(r_i)}}{1.5 + 1.4142/\sqrt{n(r_i)}}, 
\end{align}
defined by SNR=0.5 in each half-dataset. However, this was found to be very conservative in Cryo-EM imaging, to which a similarity may be drawn in CT imaging. Instead, \cite{van_heel_fourier_2005} proposes a half-bit threshold as
\begin{align}
     T_{\text{1/2-bit}}(r_i) &= \frac{0.2071 + 1.9102/\sqrt{n(r_i)}}{1.2071 + 0.9102\sqrt{n(r_i)}},\label{eq: halfbit}
\end{align}
which marks a SNR value of 0.2071 in each half image, as demonstrated in Figure \ref{fig: PC FRC}. Resolution measurements in section \ref{sec: Low dose studies} hence use Equation \eqref{eq: halfbit} as the defining threshold, averaging over 10 slices.    

\subsection{Dosimetry}
\label{sec: dosimetry}
To quantify the radiation dose associated with the images acquired herein, we utilised a PTW Farmer Ionization Chamber 30010 to measure incident flux via air KERMA. Air Kerma relates to incident photon flux $\phi_i$ through
\begin{align}
    K &= \phi_i E \mu_{tr}
\end{align}
where $K$ is KERMA, $E$ is the X-ray energy and $\mu_{tr}$ is the mass-energy transfer coefficient of the relevant material. Under negligible magnification, detector quantum efficiency $\epsilon$ can then be calculated through 
\begin{align}
    \epsilon &= \frac{\phi_D}{\phi_{i}} = \frac{\phi_D E \mu_{tr}}{K} \label{eq: epsilon}
\end{align}
where $\phi_D$ is the photon flux measured by the imaging detector and $\phi_{i}$ is the incident flux measured via air KERMA. Applying Eq. \eqref{eq: epsilon} to the imaging parameters used during the low-flux experiments in section \ref{sec: Low dose studies} produces the associated quantum efficiencies required to scale measurements for absolute dosimetry. For \unit{40}{keV} the quantum efficiency was $0.66$, while \unit{50}{keV} and \unit{60}{keV} share an efficiency of $0.65$. These values are reduced in comparison to characterisation studies performed with a similar detector model \cite{fardin_characterization_2023} and so may lead to a conservative estimate of dose. 
\begin{figure}[h!]
	\centering
	\begin{overpic}[width=0.60\columnwidth]{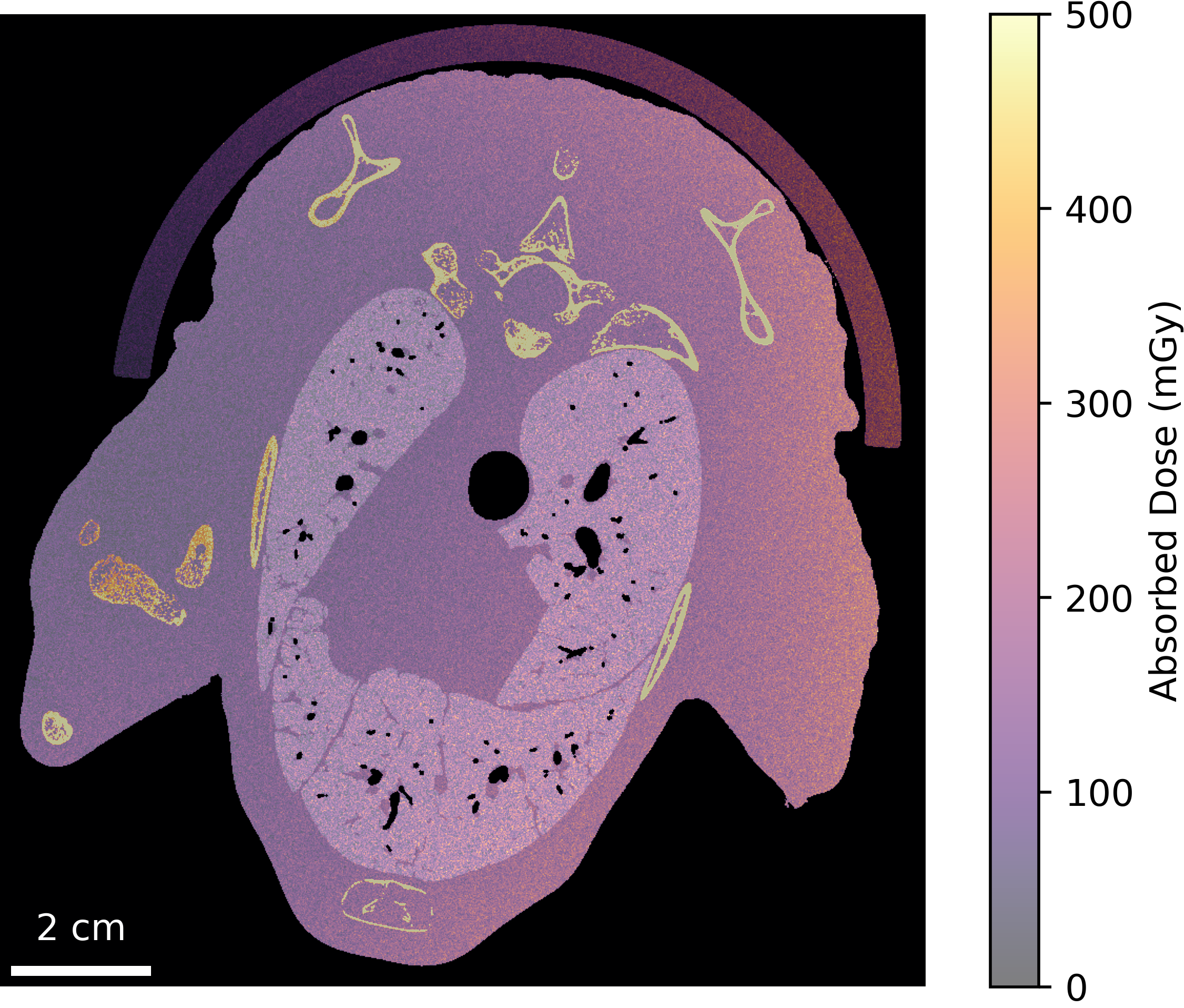}
    \end{overpic}
	\caption{Demonstration of the Geant4 dosimetry simulation using a high-flux dataset recorded at \unit{40}{keV} and a \unit{3}{m} propagation distance. The slice shown depicts the material segmented model used in Monte Carlo simulations, rebinned by $\times2$ in the $x$ and $y$ axes to reduce computation times, overlaid with the resulting dosimetry heat map. The material designations, in order of decreasing opacity, are air (black), PMMA, flesh, lung tissue and bone (white). Dose values on the right side are higher than the left because CT projections were recorded over 180 degrees. }
	\label{fig: Dose example}
\end{figure}

Absorbed dose throughout the object was determined through Monte Carlo simulation using the Geant4 toolkit. Simulations began by constructing a voxelised representation of the sample geometry. Low-noise CT reconstructions were segmented into distinct material types; PMMA for the sample holder, ICRP cortical bone for the skeletal structure, and water for an analogous material to flesh. Lung tissue was represented as a $31\%$-by-volume combination of water and the remainder air, as determined from the mean linear attenuation coefficient, $\mu$, over the lung region of the CT slices. During segmentation, each slice was rebinned in $x$ and $y$ by a factor of two for faster computations while still allowing slice-by-slice comparison. Photon trajectories were simulated at a maximum stepsize of \unit{10}{\mu m} with the G4EmPenelopePhysics physics list, using monochromatic energy incident from a randomly sampled projection angle. This produced a dose map across each slice of the CT stack, as demonstrated in Figure \ref{fig: Dose example}. Initially, the incident wavefield was weighted according to the beam profile, however scatter from high-flux regions dominated off-peak dose determinations. Therefore, a uniform beam profile was used to model the dose that would be seen under uniform-illumination clinical conditions.

To avoid scattering contributions being underestimated at the top and bottom edges of the sample, representative dosimetry values for each material were averaged from a 30-slice reference region in the middle of the beam. These values were then applied to the slices being analysed, creating a weighted-mean dose based on the amount of each material present, excluding the PMMA sample holder (the semi-circle seen in Fig. \ref{fig: Dose example}). Finally, a correction factor of $1/\epsilon$ was applied to account for the measured detector inefficiency at each energy. 

\newpage
\section{Experimental Optimisation}
\label{sec: Experimental optimisation}
Herein, we present the results of the phase contrast CT parameter optimisation for propagation distance and energy. 

\subsection{Propagation Distance Optimisation}
From Equation \ref{equation: paganin}, increasing the sample-to-detector propagation distance increases the blurring width applied during phase retrieval, counteracting the stronger Fresnel fringes produced. Hence, in this near-field regime, SNR continuously increases with propagation distance without producing a maxima \cite{kitchen_ct_2017}. However, spatial resolution may be compromised if the phase-contrast fringe strength is not in proportion with the filtering applied, for example due to penumbral blurring at large propagation distances. Therefore, optimising propagation distance is typically performed by quantifying spatial resolution through edge blur across a well-defined, high-contrast material boundary. For lung imaging, the resolution could be measured this way using large airways like the trachea, but that would disregard features of the smallest scale such as minor airways and alveoli. Such regions provide lower contrast but possess a range of feature sizes, populating a large proportion of spatial frequencies in the image. FRC$_{\text{or}}$ is ideal for analysing disperse structures like the lung but relies on the presence of noise to assess feature visibility and hence loses validity in low-noise datasets. Although the propagation optimisation datasets were recorded at high image flux, the phase contrast CTs incorporate sufficient noise to apply FRC$_{\text{or}}$. Here, phase contrast fringes amplify the sample signal within many frequency bins, thereby increasing correlation and the resulting FRC$_{\text{or}}$ spatial resolution. Therefore, in Figure \ref{fig: PC FRC}, increasing propagation distance for the unfiltered images shows increasing FRC$_{\text{or}}$ spatial resolution before plateauing after \unit{4}{m}, where penumbral blurring begins to limit development of the contrast fringes. This presents \unit{4}{m} as the optimum propagation distance, suggesting that further increases no longer enhance spatial resolution in proportion to the subsequent phase retrieval filter strength.
\begin{figure}[b!]
	\centering\hfil
	\begin{overpic}[width=0.24\columnwidth,,]{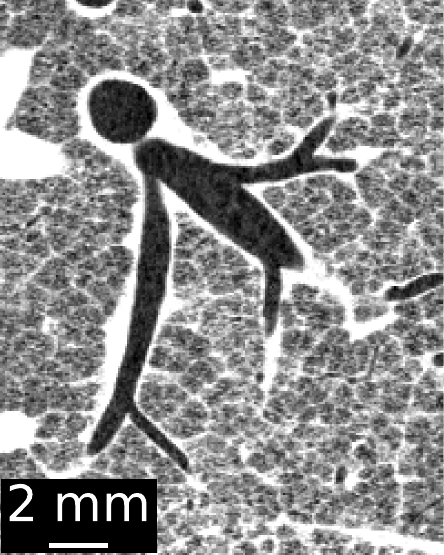}
    \put(2.5,87.5){\shortstack[l]{\fontsize{22}{24}\selectfont \textcolor{black}{a)}}}
    \end{overpic}\hspace{0.5em}
	\begin{overpic}[width=0.24\columnwidth,,]{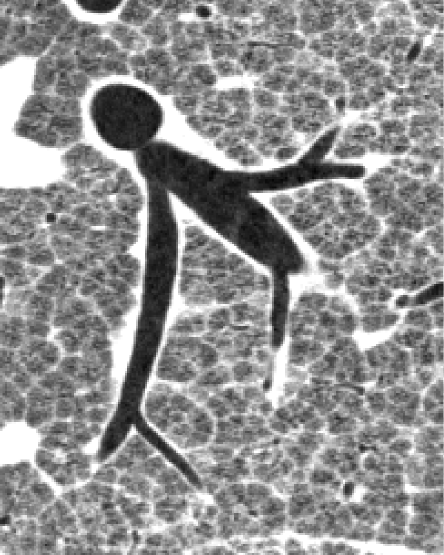}
    \put(2.5,87.5){\shortstack[l]{\fontsize{22}{24}\selectfont \textcolor{black}{b)}}}
    \end{overpic}\hspace{0.5em}
	\begin{overpic}[width=0.24\columnwidth,,]{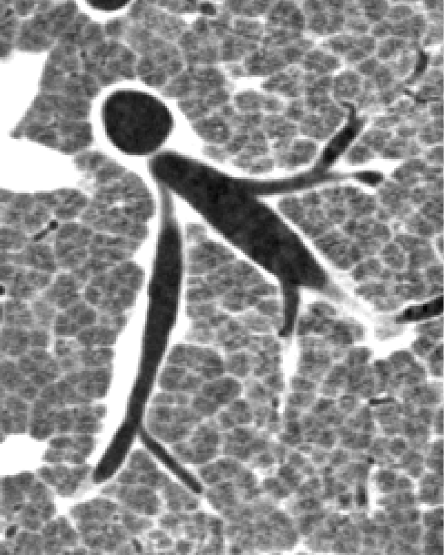}
    \put(2.5,87.5){\shortstack[l]{\fontsize{22}{24}\selectfont \textcolor{black}{c)}}}
    \end{overpic}
    \hfil
    \vspace{0.25em}
    \begin{overpic}[width=0.55\columnwidth,,]{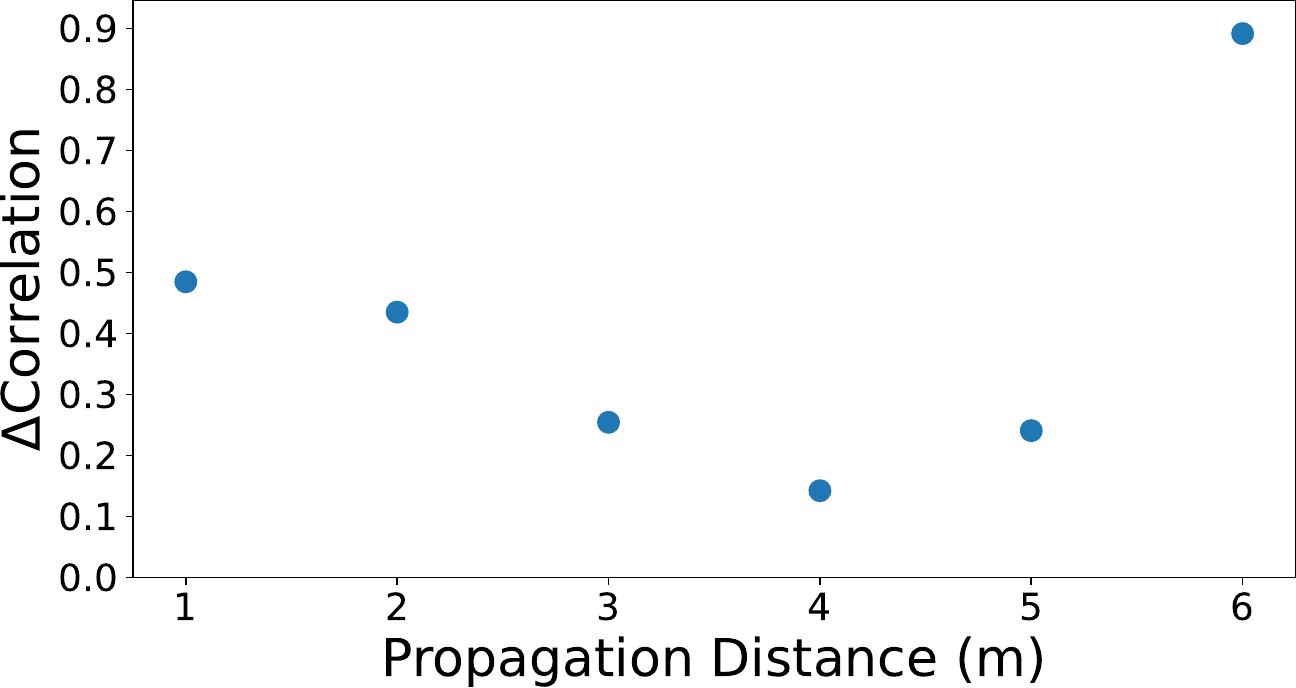}
    \put(-3.,47.5){\shortstack[l]{\fontsize{22}{24}\selectfont \textcolor{black}{d)}}}
    \end{overpic}
	\caption{Optimisation of propagation distance against resolution under constant dose. Example segments of the phase retrieved CTs are shown for propagation distances; (a) 2m, (b) 4m, and (c) 6m. (d) shows the optimisation graph, plotting the difference in mean correlation between the phase contrast and phase retrieval half-datasets, providing a visible minima at 4m.}
	\label{figure: Propdistance}
\end{figure} 
\begin{figure}[hb!]
    \centering
	\begin{overpic}[width=0.70\textwidth,,]{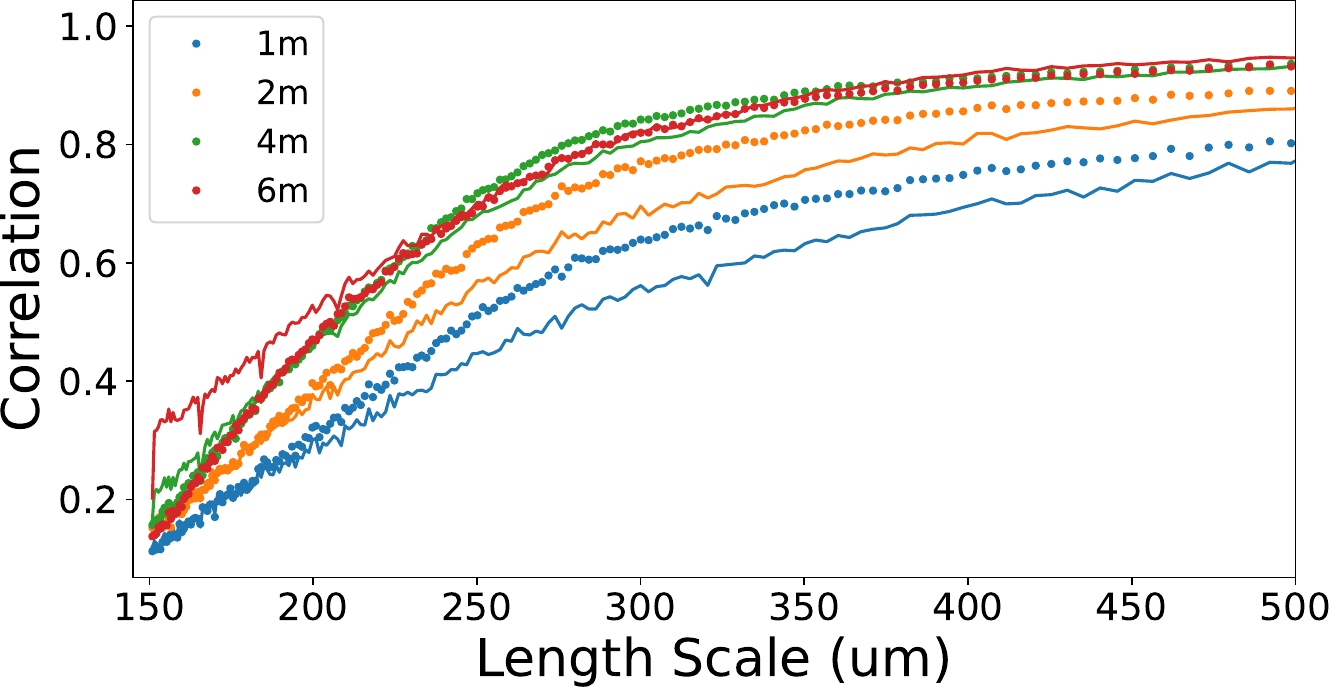}
    \end{overpic}
    \caption{Trends comparing the effect of phase retrieval on Fourier ring correlation at varying propagation distances from the sample to detector. Solid lines denote FRC$_{\text{or}}$ trends with phase retrieval applied prior to CT reconstruction, while dotted lines use CT data reconstructed from the phase contrast data.}
	\label{fig: PR FRC}
\end{figure}

Applying phase retrieval to the propagation distance data, recorded at \unit{45}{keV} and high dose, produces low-noise images as shown in Figures \ref{figure: Propdistance}(a-c). At low noise the behaviour of FRC$_{\text{or}}$ changes and the threshold intercept method outlined in Section \ref{sec:FRC} is no longer appropriate, prompting the need for a modified approach. While FRC$_{\text{or}}$ is insensitive to post-processing blurring (Section \ref{sec:FRC}, Equation \eqref{eq:FRC}), this is only observed when phase retrieval is applied directly to the CT volumes and not necessarily when applied to projections prior to CT reconstruction, due to their nonlinear nature. Figure \ref{fig: PR FRC} compares phase-contrast and phase-retrieved FRC$_{\text{or}}$ curves, showing that the FRC$_{\text{or}}$ curve changes significantly when phase retrieval is applied to data collected at both small and large propagation distances. At intermediate distances the phase-retrieved and phase-contrast trends align, satisfying the expected behaviour of FRC$_{\text{or}}$ under post-processing filtering. Optimising for the minimal difference between phase contrast and phase retrieved trends using least-squares produces the graph in Figure \ref{figure: Propdistance}(d), also showing a minima at \unit{4}{m}. Visual comparisons of the examples in Figures \ref{figure: Propdistance}(a-c) correlate with this result, with \ref{figure: Propdistance}(a) showing higher levels of noise (at \unit{2}{m}) and \ref{figure: Propdistance}(c) appearing over-blurred (at \unit{6}{m}). 

\newpage
\subsection{Energy Optimisation}
\label{sec: energy optimisation}
Optimising X-ray energy primarily concerns $\text{SNR}/\sqrt{\text{Dose}}$ as the spatial resolution should not change significantly within the energy range used here with the Eiger detector. Examples of intrinsic image quality ($Q$) trends are shown for consistent photon fluxes at each energy, however further analysis of resolution is addressed within the low flux regime in Section \ref{sec: Low dose studies}. CTs were recorded using a conservative propagation distance of \unit{3}{m}, to provide a good balance between phase contrast and penumbral blurring, at energies of 40, 45, 50, and 55 keV. 

\begin{figure}[h!]
	\centering 
    \begin{overpic}[width=0.70\columnwidth]{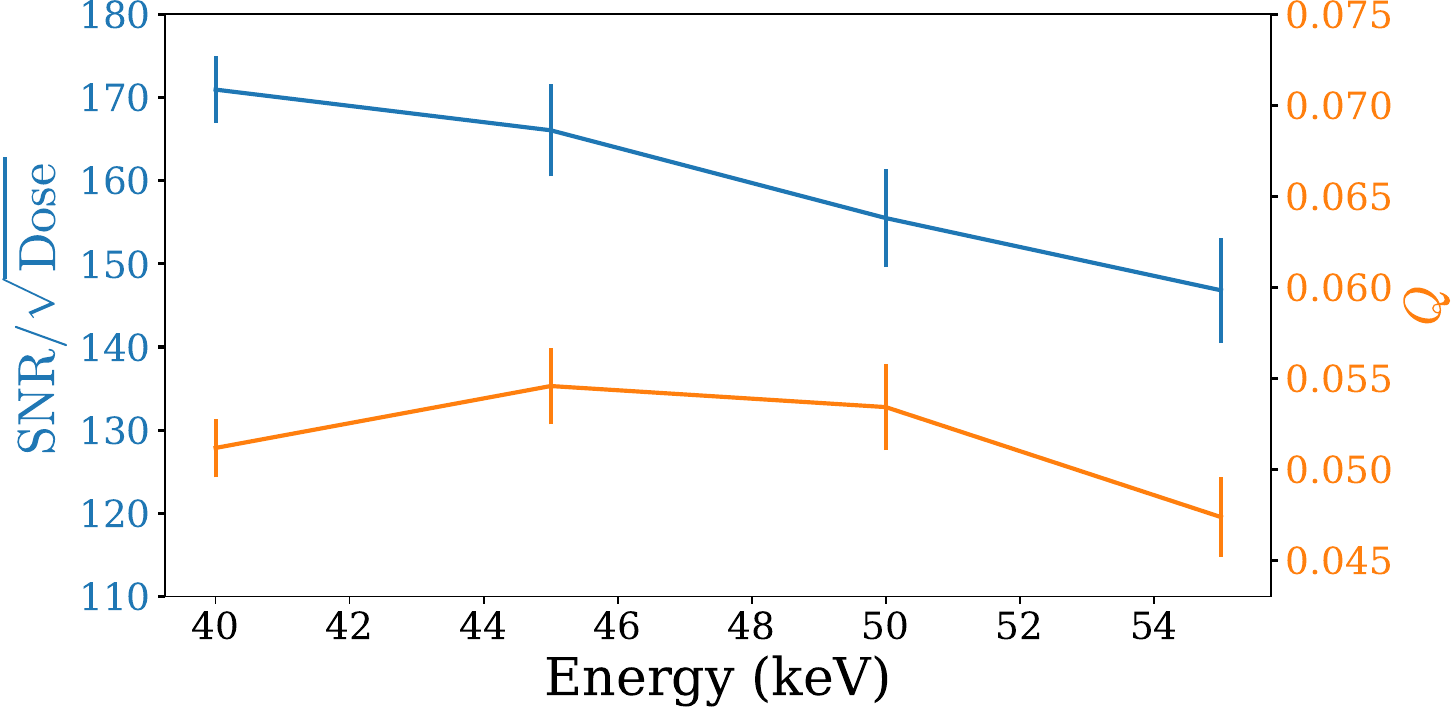}
    \put(0,42.5){\shortstack[l]{\fontsize{22}{24}\selectfont \textcolor{black}{a)}}}
    \end{overpic}\vspace{0.5em}%
    \hfil
    \begin{overpic}[width=0.35\columnwidth,,]{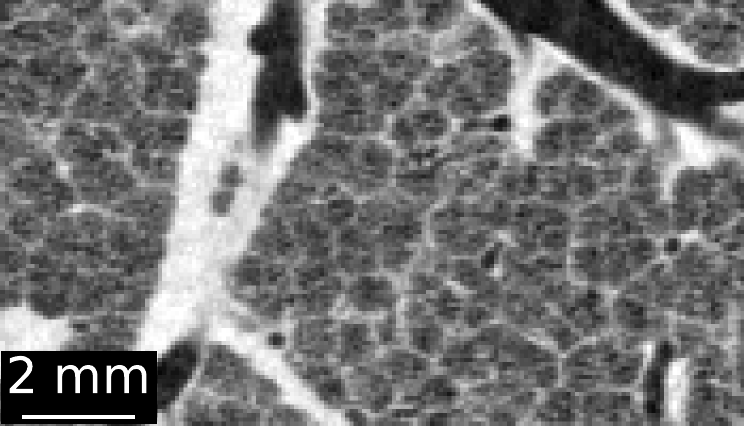}
    \put(2,45){\shortstack[l]{\fontsize{22}{24}\selectfont \textcolor{white}{b)}}}
    \end{overpic}\hspace{0.5em}%
    \begin{overpic}[width=0.35\columnwidth,,]{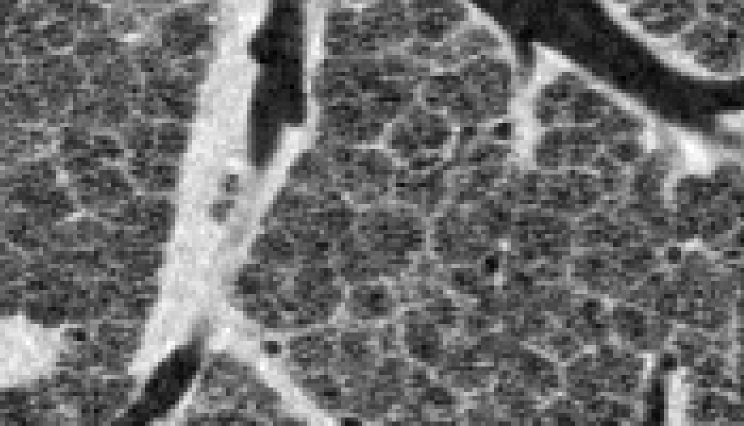}
    \put(2,45){\shortstack[l]{\fontsize{22}{24}\selectfont \textcolor{white}{c)}}}
    \end{overpic}\hfil%
    \caption{Energy optimisation of image quality according to Equation \eqref{eq: SNRperDose} and \eqref{eq: SNR/dose/res}. (a) shows a dual-axis optimisation graph for $\text{SNR}/\sqrt{\text{Dose}}$ and $Q$ at a common photon flux of 480 in the flat field. Both trends indicate 45 keV as the optimum of the energies tested. Example CT slices are also shown for (b) the optimal energy of \unit{45}{keV} and (c) the least optimal of \unit{55}{keV}, mainly due to contrast reduction.}
	\label{fig: energy}
\end{figure}

Figure \ref{fig: energy}(a) displays a figure of merit optimisation graph calculated using Equations \eqref{eq: SNRperDose} and \eqref{eq: SNR/dose/res}. Scaling of the dosimetry results with energy was performed by dividing by the theoretical absorption that would occur in the \unit{0.75}{mm} deep Cadmium Telluride pixels. Each energy uses a different CT slice to ensure equivalent flux despite variance of the synchrotron beam profile with energy. The $\text{SNR}/\sqrt{\text{Dose}}$ appears to linearly decrease with energy, whereas incorporation of FRC$_{\text{or}}$ measurements indicates \unit{45}{keV} as the optimum energy for intrinsic image quality, $Q$. Figure \ref{fig: energy}(b) shows a CT cross-section at the optimal energy of \unit{45}{keV} and \ref{fig: energy}(c) shows the least optimal of the tested energies at \unit{55}{keV}. The $Q$ graph is fairly flat between \unit{45}{keV} and \unit{50}{keV}, suggesting an intermediate optimum or a wide tolerance range for the optimisation. However, given FRC$_{\text{or}}$ measures the visibility of features as predominately affected by noise, optimisation of $Q$ may change as flux drops and hence is further explored during low dose imaging in Section \ref{sec: Low dose studies}.

\section{Low Dose Imaging}
\label{sec: Low dose studies}
A propagation distance of 4m was chosen for the low-flux studies which, from Section \ref{sec: Experimental optimisation}, aligns with the optimum propagation distance. Energies of 40, 50 and \unit{60}{keV} were used to explore the effect of increasing sample penetration on FoM qualities at low-flux. In particular, the reliability of distinguishing sample features from noise is quantified using FRC$_{\text{or}}$, representing a combined metric for the effects of contrast, scatter, and resolution. 
\begin{figure}[htb!]
    \centering 
	\begin{overpic}[width=0.49\textwidth, angle = 0,,]{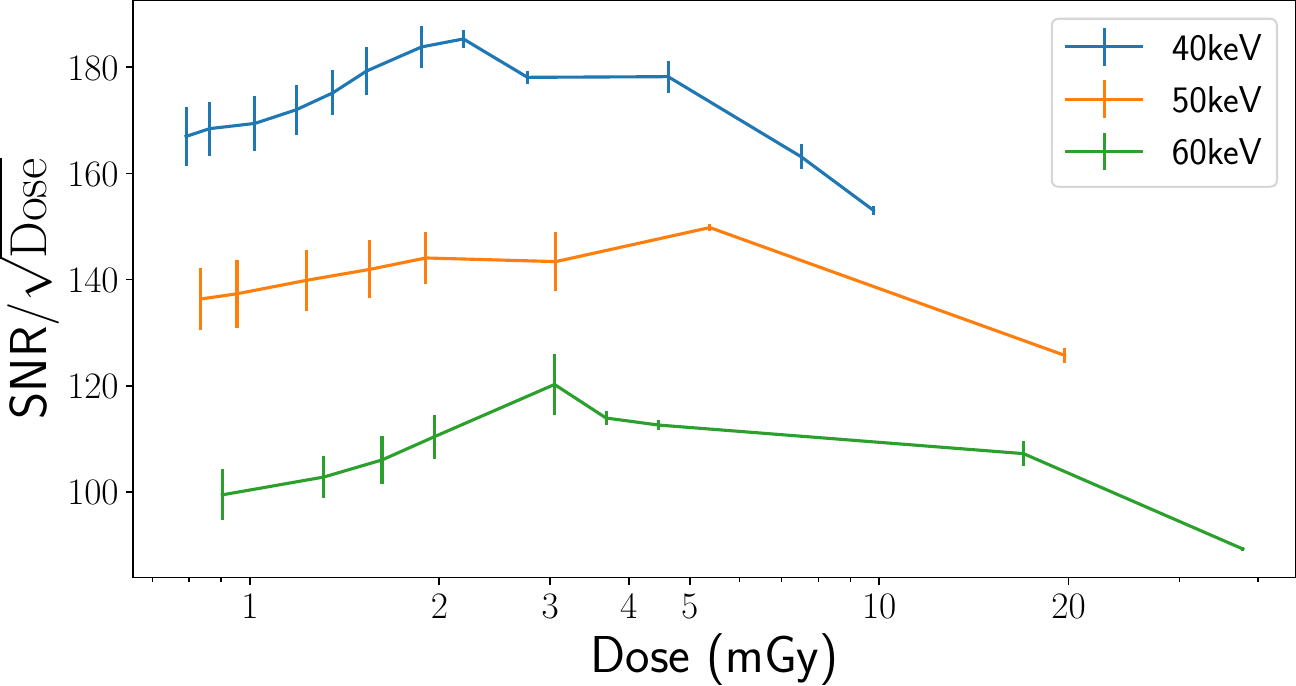}
     \put(-1,48){\shortstack[l]{\fontsize{16}{18}\selectfont \textcolor{black}{a)}}}
    \end{overpic}\hspace{0.5em}%
	\begin{overpic}[width=0.49\textwidth, angle = 0,,]{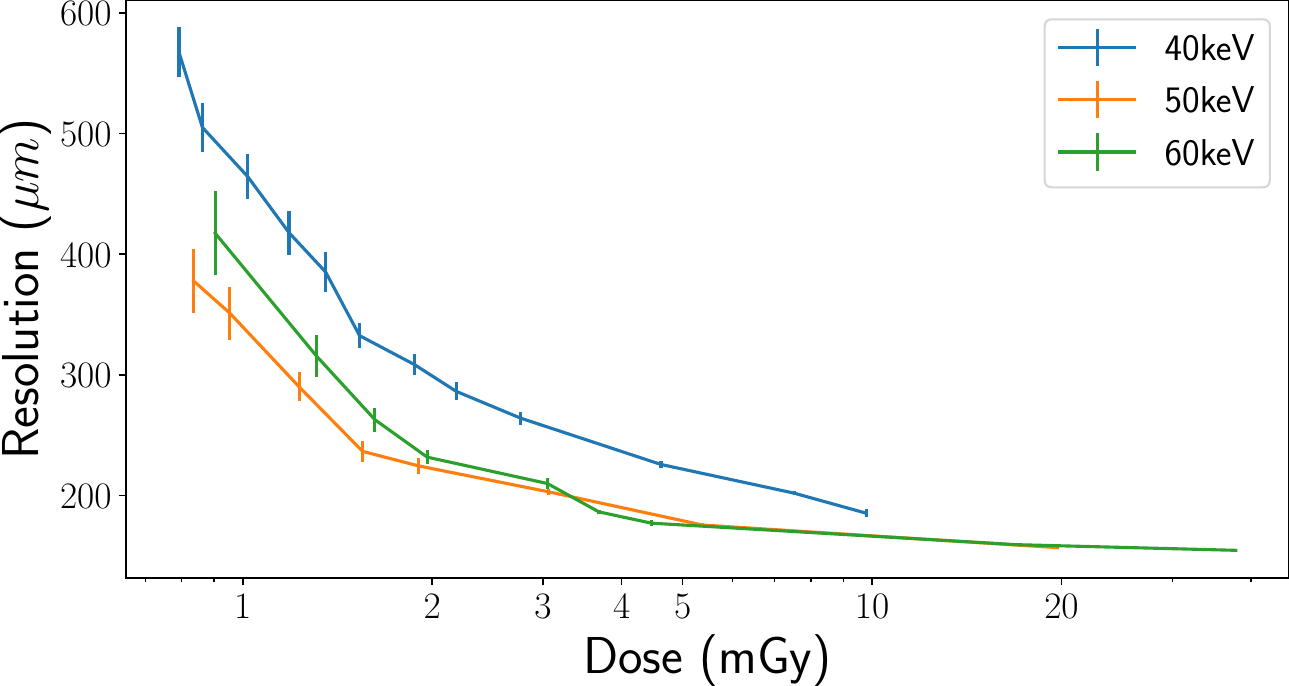}
    \put(-1,48){\shortstack[l]{\fontsize{16}{18}\selectfont \textcolor{black}{b)}}}
    \end{overpic}
    \begin{overpic}[width=0.49\textwidth, angle = 0,,]{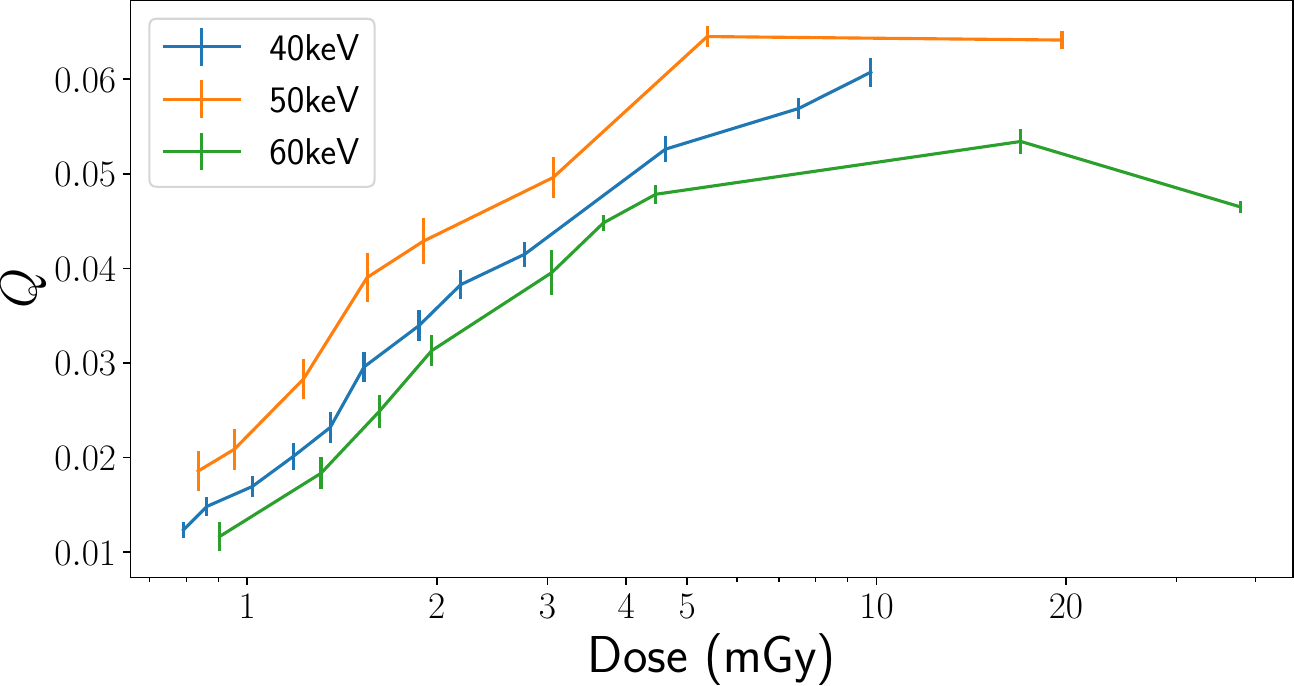}
    \put(-1,48){\shortstack[l]{\fontsize{16}{18}\selectfont \textcolor{black}{c)}}}
    \end{overpic}\hspace{0.5em}%
	\caption{Log scale figure of merit graphs comparing low-dose CT reconstructions recorded at 40, 50, and \unit{60}{keV}. (a) $\text{SNR}/\sqrt{\text{Dose}}$ according to Equation \eqref{eq: SNRperDose}, indicating \unit{40}{keV} as the optimum of the three energies. (b) FRC$_{\text{or}}$ resolution measured against dose, demonstrating resolution deteriorating as count rate drops. Here the \unit{40}{keV} trend is shown to have the worst resolution, likely owing to decreased penetration at lower energies and increased scatter. These resolution values were then used to calculate the $Q$ factors in (c). From this we see the \unit{50}{keV} trend has the optimal $Q$ at all doses. }
	\label{fig: lowdose graphs}
\end{figure}

Plots describing how image quality changes with dose are presented in Figure \ref{fig: lowdose graphs}. While $\text{SNR}/\sqrt{\text{Dose}}$ (Figure \ref{fig: lowdose graphs}(a)) is highest for \unit{40}{keV}, spatial resolution measured through FRC$_{\text{or}}$ is shown to be worst at this energy (Figure \ref{fig: lowdose graphs}(b)). This is likely due to the lower energy leading to increased absorption and scatter by the sample decreasing the SNR and hence the FRC$_{\text{or}}$ resolution (larger values = lower resolution). A sharp increase in per-photon radiation dose is also observed at \unit{40}{keV}, which correlates with increased sample absorption and scatter. Note that resolution in the context of FRC$_{\text{or}}$ does not represent a blurring limit for resolution, but rather the minimum feature size that can be consistently resolved from noise, and therefore will be closely related to the number of photons transmitted through the sample. Figure \ref{fig: lowdose graphs}(c) plots the intrinsic quality factor, $Q$, in which \unit{50}{keV} produces the clearest image in the low-dose regime, aligning with visual inspection. Per Section \ref{sec: energy optimisation}, the true maxima may lie between 45 and \unit{50}{keV}, however this was not known during data collection. Intrinsic image quality is reduced at low dose, due to the increasing contribution of Poisson noise in obscuring small features, but steadily increases with dose before plateauing above 5mGy as FRC resolution stabilises and further SNR improvement has diminishing return in comparison to the dose required.

\begin{figure}[htb!]
    \hspace{0.5cm}
    \begin{minipage}{0.530\textwidth}%
        \begin{overpic}[height=\textwidth, angle = -90,,]{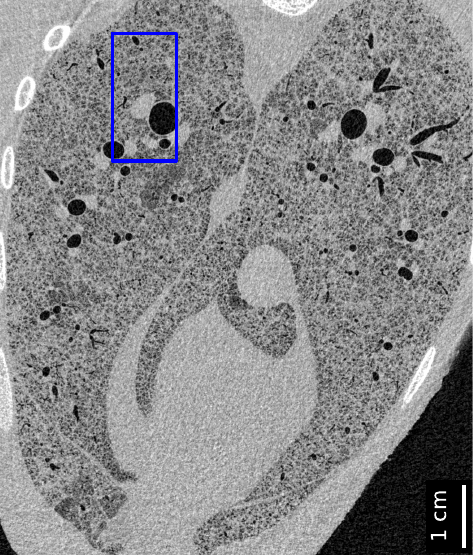}
        \put(1,78.0){\shortstack[l]{\fontsize{22}{24}\selectfont \textcolor{white}{a)}}}
        \end{overpic}
    \end{minipage}%
    \hspace{0.5mm}
    \begin{minipage}{\textwidth}%
        \begin{minipage}{\textwidth}%
            \begin{overpic}[height=0.45\textwidth, angle = -90,,]{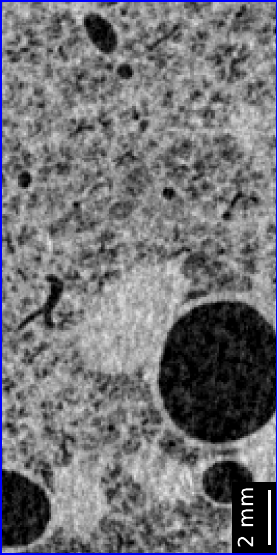}
            \put(1,41.5){\shortstack[l]{\fontsize{22}{24}\selectfont \textcolor{white}{b)}}}
            \end{overpic}
        \end{minipage}
            \begin{overpic}[height=0.45\textwidth, angle = -90,,]{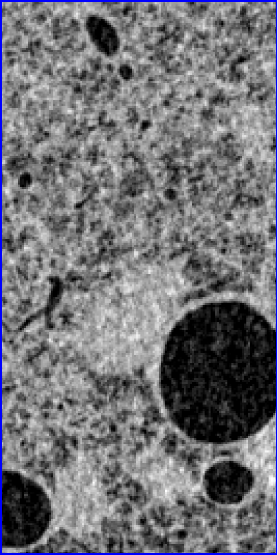}
            \put(1,41.5){\shortstack[l]{\fontsize{22}{24}\selectfont \textcolor{white}{c)}}}
            \end{overpic}
    \end{minipage}
    \caption{Example CT slices recorded with a 4 m propagation distance using \unit{50}{keV} monochromatic radiation. (a) shows a slice recorded using \unit{3.06}{mGy} absorbed dose with magnified blue inset (b) to demonstrate effectiveness in resolving minor airway features. (c) The same magnified region at a lower dose, \unit{1.91}{mGy}. Application of FRC$_{\text{or}}$ determines the spatial resolution to be (a) \unit{205}{\mu}m and (c) \unit{221}{\mu}m.}
    \label{fig: 40keV Example1}
\end{figure}

\begin{figure}[htb!]
    \hspace{0.5cm}
    \begin{minipage}{0.53\textwidth}%
        \begin{overpic}[height=\textwidth, angle = -90,,]{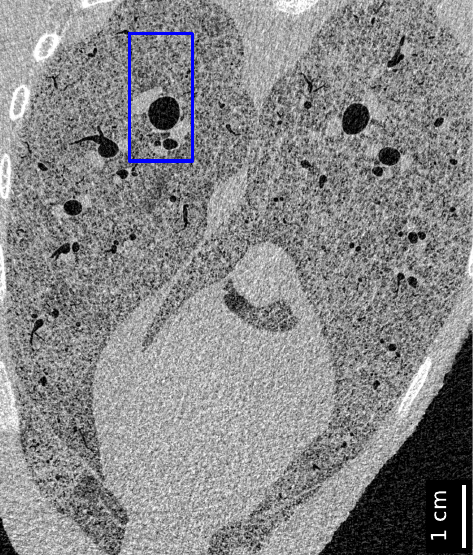}
        \put(1,78.0){\shortstack[l]{\fontsize{22}{24}\selectfont \textcolor{white}{a)}}}
        \end{overpic}
    \end{minipage}%
    \hspace{0.5mm}
    \begin{minipage}{\textwidth}%
        \begin{minipage}{\textwidth}%
            \begin{overpic}[height=0.45\textwidth, angle = -90,,]{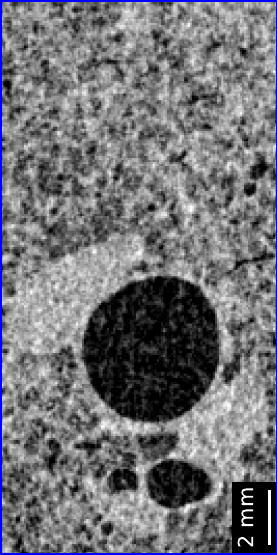}
            \put(1,41.5){\shortstack[l]{\fontsize{22}{24}\selectfont \textcolor{white}{b)}}}
            \end{overpic}
        \end{minipage}
            \begin{overpic}[height=0.45\textwidth, angle = -90,,]{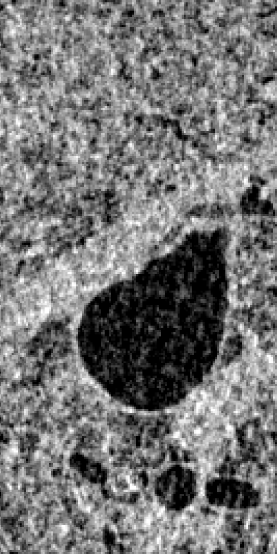}
            \put(1,41.5){\shortstack[l]{\fontsize{22}{24}\selectfont \textcolor{white}{c)}}}
            \end{overpic}
    \end{minipage}
    \caption{Example CT slices recorded with a 4m propagation distance using \unit{50}{keV} monochromatic radiation. (a) shows a slice recorded using \unit{1.22}{mGy} absorbed dose with magnified blue inset (b) to demonstrate effectiveness in resolving minor airway features. (c) shows a similar magnified region at a lower dose, \unit{0.84}{mGy}. Application of FRC$_{\text{or}}$ determines the spatial resolution to be (a) \unit{290}{\mu}m and (c) \unit{407}{\mu}m. }
    \label{fig: 40keV Example2}
\end{figure}

\begin{figure}[t!]
	\centering
	\begin{overpic}[width=0.70\textwidth,]{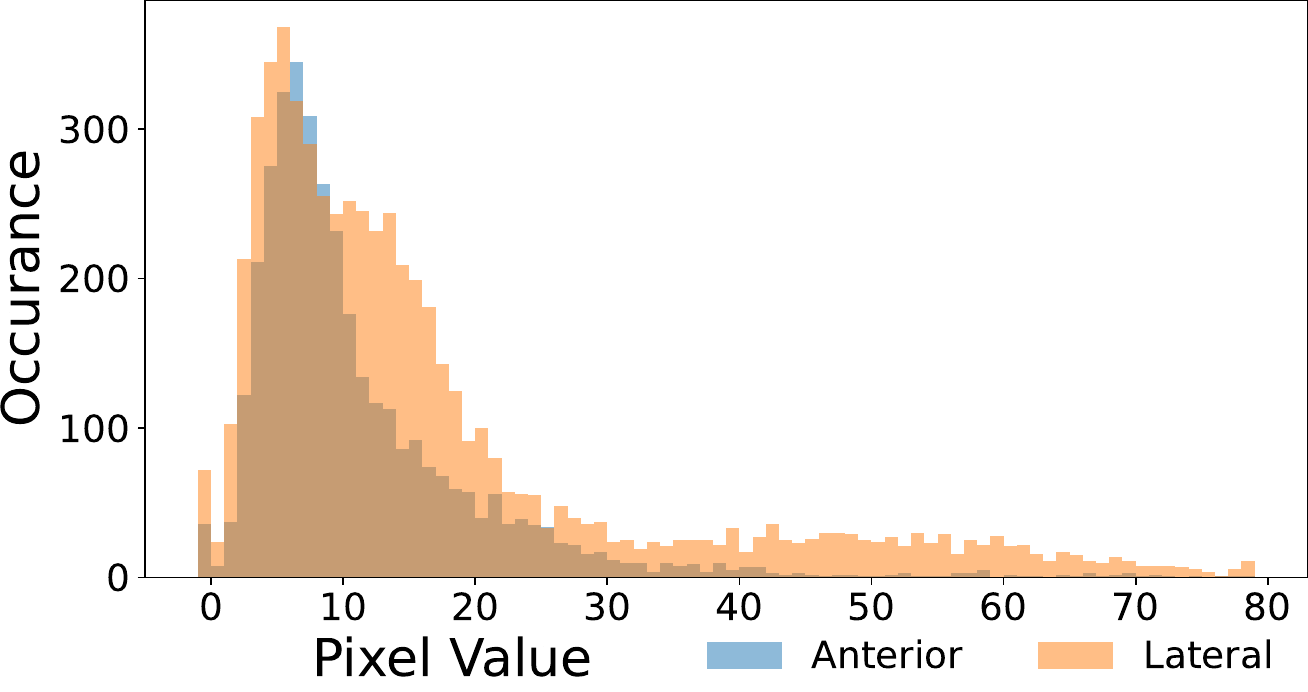}
     \put(53,52.33){
     \includegraphics[scale=0.601, angle =-90]{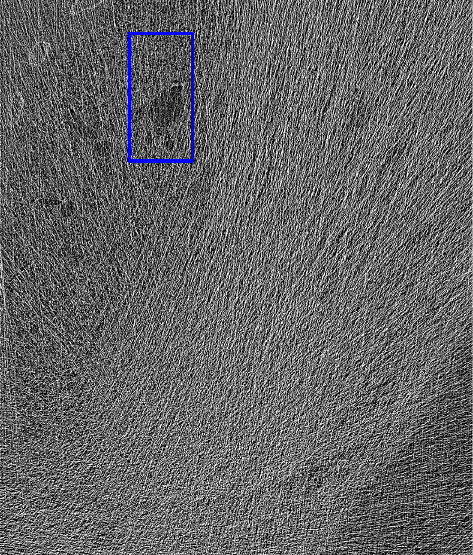}}
    \end{overpic}
    \caption{Histogram of the raw data counts recorded through the object for anterior (front) and lateral (side) views at \unit{0.84}{mGy} absorbed dose, corresponding to projections used to reconstruct the CT slice shown in Figure \ref{fig: 40keV Example2}(c). A large number of pixels showed counts of -1, denoting dead pixels and chip boundaries, with many pixels having from 0 to 1 counts, showing the extremely low dose. The inset shows the resulting CT reconstruction without application of phase retrieval, which possesses little to no diagnostic value. }
	\label{fig: Histogram}
\end{figure}

After using FoM to establish 50 keV as the ideal energy from the low-flux datasets tested, visual inspection is required to determine the minimum dose that still produces suitable image quality. Figure \ref{fig: 40keV Example1} shows comparisons of CT slices at \unit{3.06}{mGy} and \unit{1.91}{mGy}, while Figure \ref{fig: 40keV Example2} provides comparisons at \unit{1.22}{mGy} and \unit{0.84}{mGy}. Although slight variations occur within each image set due to the sample shifting between exposures, Figures \ref{fig: 40keV Example1}(b) and (c) both provide clear definitions of alveolar clusters and minor airways. Similarly, Figure \ref{fig: 40keV Example2}(b) can still resolve small sample features in the zoomed insets, while \ref{fig: 40keV Example2}(c) noticeably reflects the decrease in resolution to \unit{407}{\mu}m, approximately \unit{5.4}{pixels}, but still demonstrates clear detail of large airways. Figure \ref{fig: Histogram} emphasises the low-dose nature of \ref{fig: 40keV Example2}(c) by plotting a histogram of the raw projection counts in anterior and lateral positions. Pixel values of zero and one are shown to be a common occurrence in each, particularly in the lateral projection due to horizontal roll-off in the synchrotron beam profile. These values must be replaced with a small non-zero number to avoid undefined values occurring when taking the log of intensity during CT reconstruction, meaning information in those pixels that have not detected X-rays is lost. Phase retrieval, by blurring across projections, effectively shares information amongst pixels to fill those containing zeros in proportion to the sample qualities, thus retaining information. An example CT reconstruction, performed without phase retrieval, is shown as an inset to Figure \ref{fig: Histogram} showing no diagnostic potential and a 60-fold decrease in SNR in comparison to the corresponding phase-retrieved slice in Figure \ref{fig: 40keV Example2}(c). Therefore, a similar quality conventional CT would require a $(60\pm15\%)^2=3600\pm31\%$ times increase in dose\cite{kitchen_ct_2017}. Similarly, the images shown in Figures \ref{fig: 40keV Example2}(a) and (b) would require a $(35\pm15\%)^2=1225\pm31\%$ times dose increase if performed without phase retrieval. Clinical CT systems instead compromise by using comparatively larger pixel sizes, demonstrating the need for innovation beyond absorption contrast techniques in order to improve imaging sensitivity whilst maintaining patient safety.
\\ \\ \\ 
\newpage
\newpage
\section{Discussion}
Current guidelines from the Australian Radiation Protection and Nuclear Safety Agency (ARPANSA) estimate effective doses for chest CTs of patients aged between 0-10 years to be between \unit{1-2.5}{mSv} \cite{noauthor_ct_nodate}. Analysis of Hirosaki University hospital scans between 2011 and 2013 for paediatric patients aged 0, 1, and 5 years indicated mean effective doses of \unit{1.2\pm0.7}{mSv}, \unit{0.9\pm0.5}{mSv}, and \unit{0.8\pm0.4}{mSv}, respectively, for chest CTs \cite{obara_estimation_2017}. However, resolution and other quality factors, which are intrinsically linked with dose, were not stated. Clinical CT scans can be performed in different modes depending on risk assessment outcomes, ranging from Ultra Low Dose CT (ULDCT), which prioritises a low dose at the expense of resolution, to Ultra High Resolution CT (UHRCT), which favours resolution at the expense of dose. Standard CT then offers a compromise between the two. Doses delivered in adult chest CTs across these modalities may vary from \unit{13}{mSv} to \unit{0.4}{mSv} \cite{suliman_low-dose_2023}. HRCT necessitates increased dose even when limited to smaller regions of an organ, as observed in paediatric brain CTs (0 to 15 years old) \cite{tap_computed_2018}. We propose PBI CT imaging as a method of removing this compromise by demonstrating its ability to produce UHRCTs within the range of the ARPANSA guidelines for dose. Results of Section \ref{sec: Low dose studies} use the mean absorbed dose throughout CT slices whereas clinical studies more commonly refer to effective dose; a derived quantity of absorbed dose weighted by the fraction of tissue being exposed and a multiplicative factor based on radio-sensitivity \cite{fisher_appropriate_2017}. Referring to the ICRP 2007 tissue weighting factors and assuming full exposure of lungs, breast and oesophagus, 75\% exposure of stomach and liver, 50\% exposure to bone-marrow and the `remainder' category, and negligible whole organ dose to the rest produces a conservative weighting factor of 0.52. If the mean slice dose is assumed to be uniform across all tissue, then the effective doses of Figures \ref{fig: 40keV Example1} and \ref{fig: 40keV Example2} become \unit{~1.59}{mSv}, \unit{~0.99}{mSv}, \unit{~0.63}{mSv} and \unit{~0.44}{mSv} respectively. This would render all images well within the ARPANSA guidelines for chest CTs. 

While dose is important for patient welfare, resolution may be interpreted as the effectiveness of an imaging system to observe fine details. Clinical CT systems resample slices on fixed image grids to ensure uniformity between scanners and improve image noise via averaging. Standard CT scanners use grid sizes of $512\times512$ while high-resolution scanners often use $1024\times1024$ or $2048\times2048$ and include post-process sharpening effects \cite{balmer_influence_2022}. Optimisations of image grid size vary between systems and applications with mixed results \cite{bartlett_high-resolution_2019, euler_1024-pixel_2020, balmer_influence_2022, miyata_influence_2020}, where larger image matrices increase noise \cite{hata_effect_2018} or even double dose \cite{zhu_improved_2017}. A theoretical voxel size of the reconstruction may be inferred by dividing the field of view by the image array size, however this only applies horizontally given slice thicknesses are often up to or greater than \unit{1}{mm} \cite{euler_1024-pixel_2020, balmer_influence_2022, zhu_improved_2017}. This presents a huge limitation on the resolution in clinical scans as object features are averaged along this thickness, even with the incorporation of photon counting detectors \cite{bartlett_high-resolution_2019}. Conventional methods for determining resolution in clinical systems typically rely on resolving gaps in high-contrast steel slit phantoms which possess an axis of symmetry parallel to the CT rotation, removing the effect of slice thickness on resolution measurements. While high-contrast objects offer a reliable measure of resolution, imaging them at high doses neglects the effects of noise on visibility. By comparison, resolution measurements performed using FRC$_{\text{or}}$ directly quantify feature clarity within lung tissue and represent the length scale at which features can be resolved from noise, making it the more appropriate metric for low-dose imaging. This makes FRC$_{\text{or}}$ a more critical measure of resolution as it more realistically evaluates an image. For example, the highest resolution currently referenced in clinical CT is \unit{0.14}{mm} when using small FOVs, large matrix sizes, high-contrast and high-flux phantoms \cite{hata_effect_2018, yanagawa_subjective_2018}. However, these measurements do not account for the non-uniform voxel size and hence are not representative of the feature resolution within the sample. Although large slice thicknesses are necessitated in clinical systems to moderate patient dose, phase retrieval allows sufficient dose reduction to remove this compromise. 

Comparisons between various forms of high and ultra-high resolution CTs have been performed using adult cadavers with slice thicknesses of \unit{0.25}{mm} (dose of \unit{19-23}{mGy CTDI_{Vol}}) \cite{yanagawa_subjective_2018} and \unit{0.50}{mm} (dose of \unit{19-23}{mGy CTDI_{Vol}}) \cite{hata_effect_2018}. Thin slice CT reconstructions of \unit{0.25}{mm} have also been performed in patients with chronic obstructive pulmonary disease (COPD) with doses of \unit{11.4}{mGy CTDI_{Vol}} \cite{tanabe_quantitative_2018}, where \unit{}{CTDI_{Vol}} is measure of absorbed dose in a standardised acrylic phantom. In comparison, no slice averaging or post-sharpening was applied to our results which use \unit{0.075}{mm} slice widths as native from the \unit{0.075\times0.075}{mm^2} detector elements used. HRCT scanners instead use detector elements of \unit{0.25\times0.25}{mm^2} or \unit{0.5\times0.5}{mm^2} but benefit from magnification factors in cone beam sources. Although FRC$_{\text{or}}$ represents a higher standard of resolution measurement, even the lowest resolution scan in Figure \ref{fig: 40keV Example2}(c) presents a higher uniform resolution than some HRCT scanning modes. Additionally, given FRC$_{\text{or}}$ denotes a noise limit to resolution, array slices could be rebinned to increase SNR, potentially maintaining or increasing resolution. This would further facilitate the low-dose benefits of PBI CT whilst still retaining competitive spatial resolution. More direct comparisons against FRC$_{\text{or}}$ resolution measurements may be found in radiological surveys that rate the ability of a system to resolve specific features from noise, for example discerning bronchi smaller than \unit{2}{mm} in diameter \cite{agostini_proposal_2020}.

The primary limitation of this study is the use of synchrotron radiation during imaging. While some medical applications have secured consistent access to synchrotron beamtimes \cite{nesterets_feasibility_2015, gunaseelan_propagation-based_2023}, such facilities are not suited for high-throughput medical applications. Hence, future experiments will explore compatibility with microfocus polychromatic sources. Although the results presented feature PBI phase retrieval as the primary means of noise suppression, other methods such as iterative reconstruction, artificial neural networks, and other post-processing techniques could be incorporated in future studies.  

\section{Conclusion}
\label{sec:conclusion}
Propagation-based phase contrast imaging, combined with photon-counting detectors, is demonstrated to provide high-resolution lung CTs at and below clinically acceptable dose levels for pediatric patients. Additionally, resolution within and between slices is shown to be improved compared to clinical HRCT scanners despite using a more critical metric in Fourier ring correlation, which accounts for feature visibility above the background noise. Optimisations and demonstrated scans use medium-sized animal models, providing progression toward safer, high-resolution imaging for humans, including infants and children. Here, doses were so low that projections contained a significant number of zero-value pixels that would normally prevent CT reconstruction or lead to information loss. Application of phase retrieval is shown to stabilise low-dose, high-resolution X-ray imaging, providing CT reconstructions with over 1000 times lower dose than conventional imaging under the same conditions, progressing research towards clinical deployment with polychromatic sources. 

\bibliography{SynchrotronLowDose}

\section*{Acknowledgements}
We acknowledge the helpful contributions of Yakov Nesterets in providing dosimetry calibrations for the Eiger detector and Tim Gureyev for insightful comments on figure of merit relationships.

\section*{Author contributions statement}
J.A.P. performed data collection, analysis and wrote the manuscript. K.M. provided intellectual insight, guidance for data analysis and edited the manuscript. L.C.P.C. assisted in data collection. E.J.P. aided in experimental preparation and performed animal handling during data collection. K.J.C. and S.B.H. aided in animal handling during data collection. C.J.H, D.H and A.M. aided in setting up experimental hardware. M.J.K. aided during data collection, provided intellectual insight and guidance for data analysis and edited the manuscript. All authors reviewed the manuscript. 

\section*{Funding}
This research was supported by the National Health and Medical Research Council (NHMRC) Ideas (Grant no. 2012443) and (Program Grant no. APP113902) as well as the Victorian Government’s Operational Infrastructure Support Program. S.H. was supported by an NHMRC Principal Research Fellowship (Grant no. APP1058537). J.A.P is supported by a Research Training Program (RTP) Scholarship and the J. L. Williams Top-up Scholarship. K.M. acknowledges support from the Australian Research Council (FT18010037). Experiments were performed at the Imaging and Medical beamline of the Australian Synchrotron, part of ANSTO (proposal M17120). 

\section*{Ethics}
This experiment used deceased lambs collected from terminal experiments.

\section*{Data Availability Statement}
The datasets used and/or analysed for this manuscript are available from the corresponding author on reasonable request.

\section*{Competing Interests}
The authors declare no competing interests.

\end{document}